\documentclass[aps,prb,preprint,superscriptaddress]{revtex4}
\usepackage{epsfig}

\begin{document}  

\title{Quantum interference in exciton-Mn spin interactions in a CdTe semiconductor quantum dot}   
\author{A. Trojnar} 
\affiliation{Institute for Microstructural Sciences,
 National Research Council, Ottawa, Canada}
\affiliation{Department of Physics, University of Ottawa, Ottawa, Canada} 
\author{M. Korkusi\'nski} 
\affiliation{Institute for Microstructural Sciences,
 National Research Council, Ottawa, Canada} 
\author{E. Kadantsev} 
\affiliation{Institute for Microstructural Sciences,
 National Research Council, Ottawa, Canada}   
\author{P. Hawrylak*} 
\affiliation{Institute for Microstructural Sciences,
 National Research Council, Ottawa, Canada} 
\affiliation{Department of Physics, University of Ottawa, Ottawa, Canada} 
\author{M. Goryca}    
\affiliation{Grenoble High Magnetic Field Laboratory, CNRS Grenoble, France} 
\author{T. Kazimierczuk}
\affiliation{Institute of Experimental Physics, University of Warsaw, Warsaw, Poland}
\author{P. Kossacki}
\affiliation{Grenoble High Magnetic Field Laboratory, CNRS Grenoble, France} 
\affiliation{Institute of Experimental Physics, University of Warsaw, Warsaw, Poland}
\author{P. Wojnar}
\affiliation{Institute of Physics, Polish Academy of Sciences, Warsaw, Poland}
\author{M. Potemski}
\affiliation{Grenoble High Magnetic Field Laboratory, CNRS Grenoble, France}

\date{\today}

\begin{abstract}
We show theoretically and experimentally the existence of a new quantum interference(QI) effect between the electron-hole interactions and the scattering by a single Mn impurity. 
Theoretical model, including electron-valence hole correlations, the short and long range exchange interaction of Mn ion with the heavy hole and with electron and anisotropy of the quantum dot, is compared with photoluminescence spectroscopy of CdTe dots with single magnetic ions. We show how
design of the electronic levels of a quantum dot enable the design of an exciton, control of the quantum interference and hence engineering of light-Mn interaction. 
\end{abstract}
\maketitle 

Isolating and controlling states of a single quantum spin either on a surface of a metal \cite{heinrich_gupta_science2004,loth_bergman_nat2010} or in a semiconductor quantum dot \cite{erwin_zu_nat2005,ochsebein_feng_NatNan2009,bussian_crooker_NatMat2009,Hundt_puls_prb2004, besomber_leger_prl2004,gall_kolodka_prb2010,goryca_kazimierczuk_prl2009,kudelski_lamaitre_prl2007} is at an early stage. 
The spin of a single Manganese (Mn) ion is an atomic limit of magnetic memory, realized recently in semiconductor quantum dots \cite{Hundt_puls_prb2004, besomber_leger_prl2004,gall_kolodka_prb2010,goryca_kazimierczuk_prl2009,kudelski_lamaitre_prl2007}. The Mn ion with magnetic moment M=5/2 has been detected by observation of a characteristic excitonic emission spectrum consisting of six emission lines related to the 2M+1=6  possible Mn quantum states. The emission spectrum has been understood based on a spin model where exciton spin interacts with the spin of the Mn ion \cite{Hundt_puls_prb2004, besomber_leger_prl2004,gall_kolodka_prb2010,goryca_kazimierczuk_prl2009,kudelski_lamaitre_prl2007, rossier_prb2006,cheng_hawrylak_prl2008,govorov_kalameitsev_prb2005,reiter_kuhn_prl2009}. However, only a microscopic treatment of an exciton as a correlated excited state of the interacting quantum dot and the Mn as an impurity allows for full control of exciton-Mn coupling. This problem is related to the nontrivial enhancement of the electron-electron interactions by impurities \cite{richardella_roushan_science2010} as well as the Kondo effect \cite{gordon_shtrikman_nat1998}.

Here we show theoretically and experimentally how one can manipulate the spin of Mn ion with light in a semiconductor quantum dot by engineering Mn-exciton interactions through design of a quantum-dot exciton \cite{bayer_stern_nat2000,hawrylak_narvaez_prl2000}. 
A new quantum interference (QI) effect between the electron-hole Coulomb scattering and the scattering by Mn ion is shown to significantly reduce the exciton-Mn coupling revealed by a characteristic pattern in the emission spectrum. Engineering light-Mn spin interaction opens up new applications in quantum memory and information processing. 

An exciton \cite{bayer_stern_nat2000,hawrylak_narvaez_prl2000} is composed of an electron with spin $\sigma=1/2$ and a valence heavy hole with spin $\tau=3/2$ occupying single-particle levels $|i\rangle=|n,m\rangle$ of two harmonic oscillators with quantum numbers $n$ and $m$ and energy $E_i$ \cite{raymond_studenikin_prl2004,hawrylak_prl1993}. The electron and hole 
%
\begin{figure}
\epsfig{file=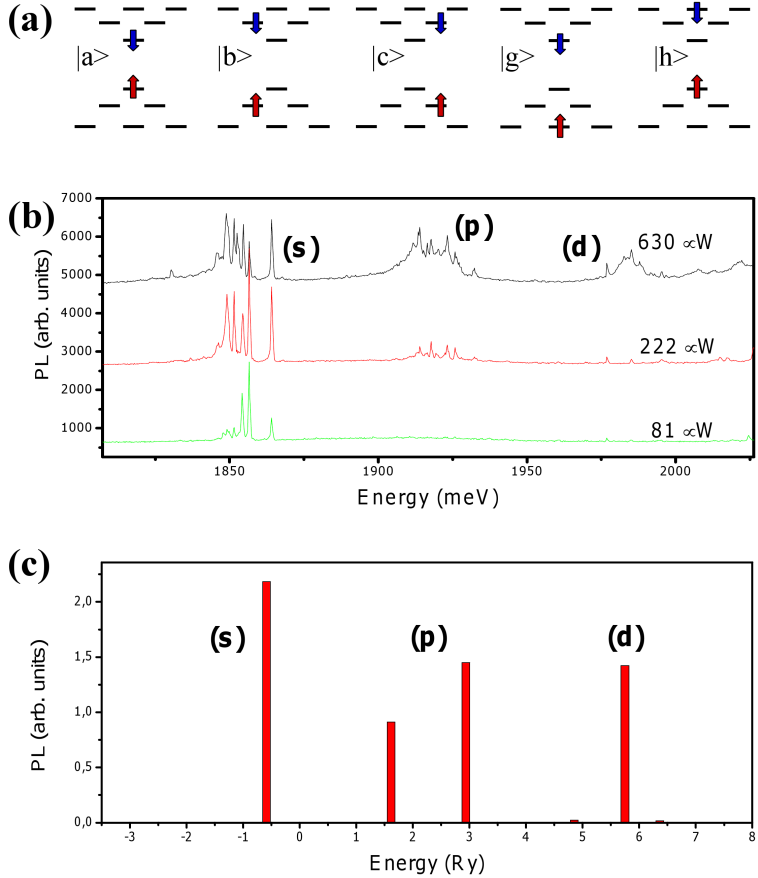,width=4in}
\caption{\label{fig_1}
(Color online) (a) The electron and hole shell structure $E_{n,m}$ and basic two-particle configurations. Electron is marked by the blue arrow, while the red arrow denotes the hole. (b) Measured photoluminescence spectra from $s$, $p$, and $d$ shells of a single CdTe quantum dot populated with increasing excitation power. (c) Calculated absorption spectrum of the CdTe isotropic quantum dot with negligibly small electron-hole exchange. Calculations were done for the single-particle energies  $\omega_e+\omega_h=30 meV$ and with $\omega_e=\omega_h$.}
\end{figure}
shell structure $E_{n,m}$ is shown in  Fig.\ref{fig_1}(a). The state of an electron-hole pair $|i,j\rangle| \sigma, \tau\rangle$ is a product of the orbital part and the spin part. The lowest energy state, labeled $|a\rangle$ in Fig.\ref{fig_1}(a), corresponds to the electron and the hole on the s shell ($n=0$, $m=0$) while excited states $|b\rangle$ and $|c\rangle$ correspond to both the electron and the hole excited from the $s$ shell to the $p$ shell ($n=0$,$m=1$; $n=1$,$m=0$).

If the $d$ shell is present in the quantum dot, another pair of excited states (labeled $|g\rangle$ and $|h\rangle$ in Fig.\ref{fig_1}(a)) at a similar energy is possible where either the hole or the electron is excited from the $s$ shell to the zero angular momentum state ($n=1$,$m=1$) of the $d$ shell. The $s$, $p$, and $d$ shells of a single CdTe quantum dot studied here appear as emission maxima with an increasing excitation power, as shown in Fig.\ref{fig_1}(b). By rotating the electron-hole configurations to Jacobi coordinates \cite{hawrylak_narvaez_prl2000} one finds that there are only three low-energy electron-hole configurations: $|A\rangle=|a\rangle$,$|B\rangle=1/\sqrt{2}\left(|b\rangle+|c\rangle\right)$ and $|H\rangle=1/\sqrt{2}\left(|h\rangle+|g\rangle\right)$coupled by Coulomb interactions. We will also refer to these configurations as $|SS\rangle$, $|PP\rangle$ and $|SD\rangle$. Only configurations $|SS\rangle$ and $|PP\rangle$ are optically active but Coulomb scattering couples all three exciton configurations, and in particular the degenerate configurations $|PP\rangle$ and $|SD\rangle$ \cite{hawrylak_narvaez_prl2000}. By diagonalizing the electron-hole Hamiltonian
 $H_{EH}=\sum_{i\tau}\varepsilon_{i\tau}^hh^+_{i\tau}h_{i\tau}+
\sum_{i\sigma}\varepsilon_{i\sigma}^ec^+_{i\sigma}c_{i\sigma}+ 
\sum_{ijkl\sigma\tau}\langle i,j|V_{eh}|k,l\rangle c^+_{i\sigma}h^+_{j\tau}h_{k\tau}c_{l\sigma}$  (where $h^+_{i\tau}$ ($c^+_{i\sigma}$) and $h_{i\tau}$ ($c_{i\sigma}$) 
are create and anihiliate  hole (electron) on the orbital $i$ with spin $\tau$($\sigma$)) in the space of all configurations 
we obtain the ground and excited states as well as the absorption spectrum, shown in Fig.\ref{fig_1}(c). We see that for a quantum dot with $s$-$d$ shells the $p$-shell splits into two lines due to the $|SD\rangle$ configuration resonant with the $|PP\rangle$ configuration \cite{hawrylak_narvaez_prl2000}, and correspondingly, contributes to the ground state $|GS\rangle$ of the exciton: $|GS\rangle=A_{ss}|SS\rangle+A_{pp}|PP\rangle-A_{sd}|SD\rangle$.
We note that the $|PP\rangle$ and $|SD\rangle$ configurations contribute to the $|GS\rangle$ with opposite signs, a result of different signs of Coulomb matrix elements $\langle SS|V|PP\rangle=-\langle SS|V|SD\rangle$ connecting the $|PP\rangle$ and $|SD\rangle$ configurations with the $|SS\rangle$ configuration.

The interacting electron-hole-Mn system is described by the Hamiltonian \cite{cheng_hawrylak_prl2008}: $H_X=H_{EH}+H_{EHX}+H_{anis}+H_{Zeeman}+H_{h-Mn}+H_{e-Mn}$. The first term is the electron-hole Hamiltonian $H_{EH}$, the second term is the electron-hole exchange term \cite{kadantsev_hawrylak_prb2010,kadantsev_hawrylak_jp2010}
$H_{EHX}=\sum_{ijkl\sigma\sigma '\tau \tau '}\langle i\sigma,j\tau|V_{eh}^X|k\tau ', l\sigma '\rangle c^+_{i\sigma}h^+_{j\tau}h_{k\tau '}c_{l\sigma '}$,
third - the anisotropic potential term 
$H_{anis}=\sum_{ij\tau}t_{ij}^hh^+_{i\tau}h_{j\tau}+
\sum_{ij\sigma}t_{ij}^ec^+_{i\sigma}c_{i\sigma}$ 
which breaks the cylindrical symmetry of the quantum dot and mixes the single particle states with different angular momenta. The fourth term is the Zeeman energy of the magnetic ion, the spin of the hole and of the electron 
$H_{Zeeman}=g_{Mn}\mu_BBM_Z+g_e\mu_B B S_Z+g_h\mu_B B J_Z$, 
where$g_e$($g_h$) are electron(hole) Lande g-factors and $\mu_B$ the Bohr magneton.
 The hole-Mn ion Hamiltonian
$H_{h-Mn}=\sum_{i,j}\frac{3J_{ij}^h\left(0\right)}{2}\left[\left(h^+_{i,\Uparrow}h_{j,\Uparrow} -h^+_{i,\Downarrow}h_{j,\Downarrow}\right)M_Z\right]$  
describes the scattering of the hole by the Mn ion while conserving the hole spin. $J_{ij}^h\left(0\right)$ is the effective exchange matrix element leading to the scattering of a hole from state $i$ to state $j$ by the Mn ion at position $R=0$ \cite{cheng_hawrylak_prl2008,qu_hawrylak_prl2005}. This scattering process does depend on the state of the Mn-ion. The electron-Mn interaction term is similar to the hole-Mn scattering term except for the additional spin flipping term $H_{e-Mn}=-\sum_{i,j}\frac{J_{ij}^e\left(0\right)}{2}\left[\left(c^+_{i,\uparrow}c_{j,\uparrow} -c^+_{i,\downarrow}c_{j,\downarrow}\right)M_Z+c^+_{i,\downarrow}c_{j,\uparrow}M^+ \right.
 \left.+c^+_{i,\uparrow}c_{j,\downarrow}M^-\right]$. 

We now turn to evaluate the exchange interaction of the exciton with the Mn spin, dominated by the valence hole-Mn Ising-like interaction \cite{cheng_hawrylak_prl2008,sheng_hawrylak_prb2006}.
The spin of the hole plays the role of the effective magnetic field, leading to the "exchange" splitting of different $M_Z$ states: $\langle H_{h-Mn}\rangle=\langle M_Z|\langle \downarrow \Uparrow|\langle GS|H_{h-Mn}|GS\rangle|\Uparrow\downarrow\rangle|M_Z\rangle=\alpha M_Z$. With $p$ orbitals not coupled to the Mn in the center of the dot \cite{qu_hawrylak_prl2005}:
\begin{equation}
\langle H_{h-Mn}\rangle=\frac{3}{2}\left[{A_{ss}^*}^2J_{ss}-\sqrt{2}A_{ss}A_{ds}J_{sd}\right]M_Z,
\end{equation} 
We see that the exchange splitting $\alpha=3/2\left[{A_{ss}^*}^2J_{ss}-\sqrt{2}A_{ss}A_{ds}J_{sd}\right]$ of Mn levels is a difference of two terms. The first term ${A_{ss}^*}^2J_{ss}$ is proportional to the product of the sum of probability amplitudes of the hole occupying $s$ and $d$ orbitals ${A_{ss}^*}^2=A_{ss}^2+A_{ds}^2$ in the exciton GS weighted by the exchange matrix element $J_{dd}=J_{ss}$. The second term, $-\sqrt{2}A_{ss}A_{ds}J_{sd}$, reduces the magnitude of the exchange. This term is proportional to the product $A_{ds}J_{sd}$, i.e., the amplitude $A_{ds}$ of the $|SD\rangle$ configuration in the exciton GS, present only due to the electron-hole Coulomb interaction, and scattering matrix element $J_{sd}$ of the hole by the Mn ion acting as an impurity. Hence both the electron-hole Coulomb interactions and the scattering by the Mn impurity must be simultaneously present to reduce the hole exchange field. This is the quantum interference (QI) effect, the central result of this work. The QI is absent in shallow quantum dots with $s$-$p$ shells but takes place in quantum dots with at least three confined shells. 

We now turn to the second signature of QI, coupling of excited exciton states with the ground state by Mn as a scattering center. The first excited state $|ES\rangle=B_{sd}|SD\rangle+B_{pp}|PP\rangle+B_{ss}|SS\rangle+\ldots$ is a linear combination of configurations $|SD>$ and $|PP>$ with a small admixture of the $|SS>$ configuration. The coupling of $|GS>$ and $|ES>$ by the hole-Mn exchange interaction $\langle M_Z|\langle \downarrow \Uparrow|\langle GS|H_{h-Mn}|ES\rangle|\Uparrow\downarrow\rangle|M_Z\rangle=\gamma M_Z$ turns out to be proportional to the state of the Mn spin $M_Z$. The excited state renormalizes the energies of the ground state exciton-Mn spin complex $E_{GS}^{M_Z}=E_{GS}+\alpha M_Z-\left(\frac{\gamma^2M_Z^2}{\left(\Delta E-\left(\beta-\alpha\right)M_Z\right)}\right)$, where  is the exchange splitting of the Mn levels in the first excited exciton state $|ES>$ with energy $E_{ES}$ and $\delta E = E_{ES} - E_{GS}$. The main result is the nonuniform and renormalized spacing of Mn energy levels in the $s$ shell:
\begin{equation}
\Delta_{M_Z}=E_{GS}^{M_Z+1}-E_{GS}^{M_Z}=\left(\alpha-\frac{\gamma^2}{\Delta E}\right)-\frac{2\gamma^2M_Z}{\Delta E}.
\end{equation}
\begin{figure}
\epsfig{file=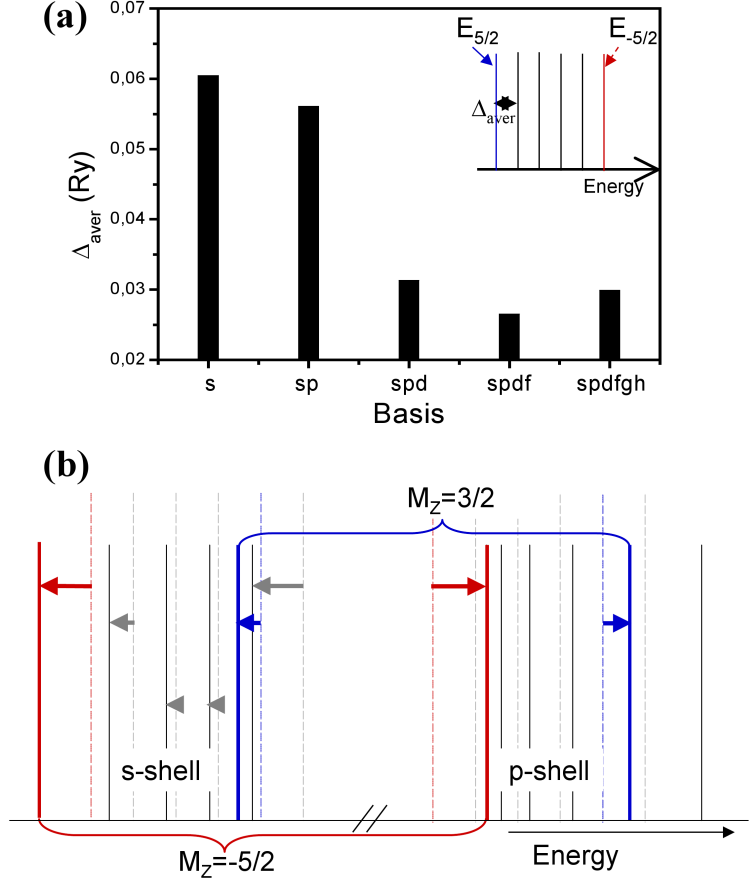,width=4in}
\caption{\label{fig_2}
(Color online) (a) Calculated average spacing ($\delta_{aver}=(E_{-5/2} - E_{5/2})/5)$ of Mn energy levels in the $s$ shell as a function of the number of shells of an isotropic CdTe quantum dot with negligible electron-hole exchange, with single-particle energies $\omega_e+\omega_h=30 meV$; $\omega_e/\omega_h=4$. (b) Schematic renormalization of $s$-shell Mn energy levels by the interaction with excited states of an exciton. Levels corresponding to the same Mn ion spin projection interact and repel each other, with the strength proportional to $M_Z$. Dashed (solid) vertical lines represent the energy levels of the X-Mn system in the $s$-shell (six lines on the left) and the $p$-shell (six lines on the right) energy region unrenormalized (renormalized) by the interaction, whose magnitude is represented by the horizontal arrows. }
\end{figure} 
Figure\ref{fig_2}(a) shows the results of numerical calculations, of the average spacing of Mn energy levels in the $s$ shell as a function of the number of shells, for parameters typical for a CdTe quantum dot. Indeed, we see that the spacing is reduced by a factor of ~2 when the quantum dot admits the $d$ shell. The renormalization of $s$-shell Mn energy levels by the excited exciton state is shown schematically in Fig.\ref{fig_2}(b). We see that the ground and excited levels corresponding to the same $M_Z$ are coupled by Mn, the coupling strength is different for each $M_Z$ leading to energy shift, with states with higher $|M_Z|$ shifting more, which in-turn leads to a nonuniform spacing of levels. The differences in the magnitude of this shift are visualized in Fig.\ref{fig_2}(b) in the form of different lengths of arrows, with the solid (dashed) vertical lines representing the exciton-Mn energy levels with (without) the ground state-excited state coupling. 

The experimental spectra of the emission from quantum dots were obtained for CdTe based heterostructures. The samples were grown using molecular beam epitaxy. Each of them contains a single layer of self-assembled CdTe QDs with a low concentration of $Mn^{2+}$ ions, embedded in a ZnTe matrix. The density of quantum dots was about $5x10^9cm^2$. The $Mn^{2+}$ concentration was adjusted to obtain a significant number of QDs containing exactly one $Mn^{2+}$ ion \cite{wojnar_suffczynski_prb2007}. For the measurements, the sample was placed in a micro-photoluminescence setup composed of piezo-electric $x-y-z$ stages and a microscope objective. The system was kept at the temperature of $4.2K$ in a helium exchange gas. The PL of the QDs was excited either above the gap of the ZnTe barrier (at $532nm$) or using a tunable dye laser in the range $570-610nm$. Both the exciting and the collected light were transmitted though a monomode fiber coupled directly to the microscope objective. The overall spatial resolution of the set-up was better then $1 \mu m$ which assured possibility to select different single quantum dots containing a single $Mn^{2+}$ ion. The dots without $Mn^{2+}$ ion were observed in the same samples. The PL analysis was done for the dots having emission lines in the low energy tail of the broad PL emission band which assured good separation from the lines related to the other dots. The characteristic PL spectra contain a neutral exciton line split into sextuplets. Lower in energy, the lines related to charged excitons ($X^+$ and $X^-$) and biexciton were observed. Higher in energy, the emission from higher shells ($s$, $p$, $d \ldots$) appear with an increasing excitation power, as is shown in Fig.\ref{fig_1}(b).
\begin{figure}
\epsfig{file=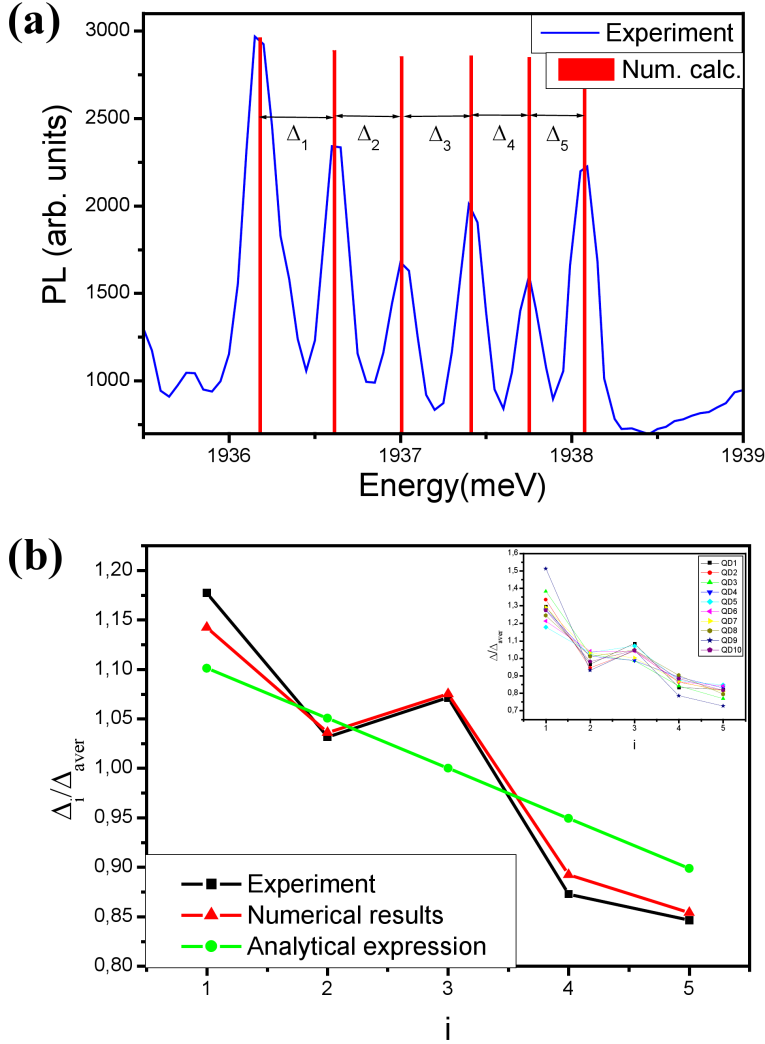,width=4in}
\caption{\label{fig_3}
(Color online) (a) Measured and calculated emission spectrum including small anisotropy ($\gamma=0.34$) and electron-hole exchange interaction $\Delta_0=0.5 meV$, $\delta_2=0.16 meV$, for the quantum dot with single-particle energies $\omega_e+\omega_h=30 meV$; $\omega_e/\omega_h=4$. (b) Comparison of the measured and calculated 
peak separation $\Delta_i / \Delta_{aver}$ ($\Delta_{aver}$ being the average distance) as a function of the peak number. The inset shows $\Delta_i / \Delta_{aver}$ aver extracted from experimental studies of ten quantum dots. The green line shows $\Delta_{M_Z} / \Delta_{aver}$  calculated analytically which neglects anisotropy and electron-hole exchange interaction.}
\end{figure}
Figure \ref{fig_3}(a) shows the measured and numerically calculated emission spectrum, including a small anisotropy of the quantum dot and the electron-hole exchange interaction
 \cite{kadantsev_hawrylak_prb2010,kadantsev_hawrylak_jp2010}. There are six emission peaks associated with $M_Z$. The predicted peak spacing $\Delta_{M_Z}$, plotted in Fig.\ref{fig_3}(b) with the green line, decreases linearly with increasing $M_Z$. This decrease is reproduced by numerical calculations and experiment (black line). Deviations from linear dependence of $\Delta_{M_Z}$  are due to the electron-hole exchange interaction and anisotropy. The inset of Fig.\ref{fig_3}(b) verifies the characteristic pattern of distances between X-Mn emission peaks for ten more different quantum dot samples.
\begin{figure}
\epsfig{file=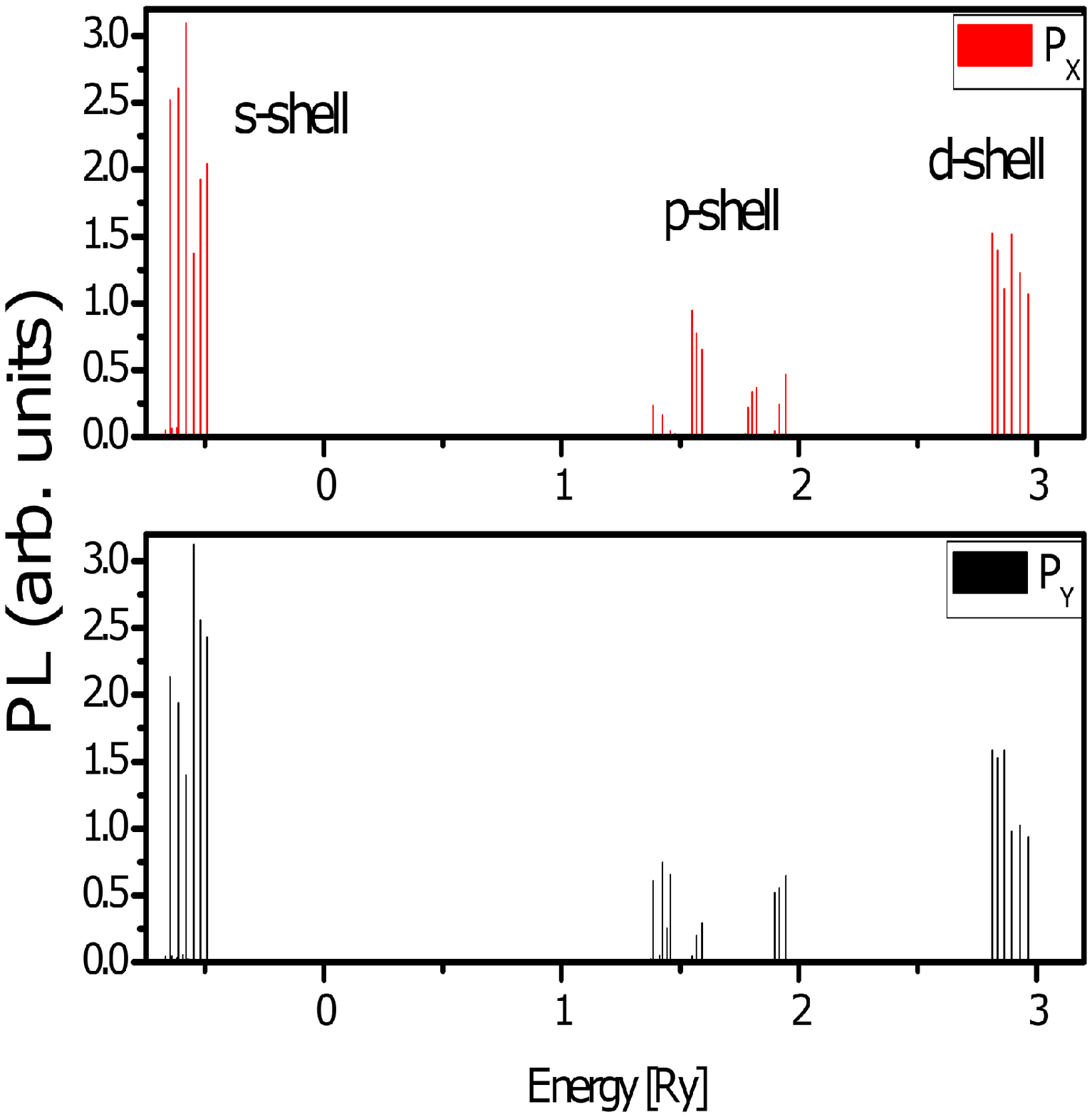,width=4in}
\caption{\label{fig_4}
(Color online) Absorption spectrum calculated for a CdTe quantum dot with parameter as on Fig. \ref{fig_3}. The color of the maxima corresponds to the degree of linear polarization of the resulting photon, with red (black) denoting $p_x$ ($p_y$) polarization.}
\end{figure}
Finally, Fig.\ref{fig_4} shows the calculated absorption spectra. We see the $s$ shell, the two excited exciton states associated with $|PP>$ and $|SD>$ configurations in the energy range of the $p$ shell, and the $d$ shell. The shells are split into a fine structure by the presence of Mn. Different colors of the peaks correspond to the degree of linear polarization of absorbed photons, with black (red) denoting the $p_y$ ($p_x$) polarization. In this spectrum we identify the two consequences of the existence of the $d$ shell: the complex emission pattern in the $p$-shell range of energies and the QI in the $s$ shell. Also, the $p$ shell experiences a much larger electron-hole exchange splitting than the $s$-shell and $d$-shell emission lines, and, in consequence, a much stronger linear polarization of the emission lines. This is due to the larger sensitivity of the $p$-shell orbitals to the shape anisotropy of the quantum dot. Experiments are on the way to verify the predicted absorption spectra.

In summary, we formulated a microscopic description of the exciton-Mn interaction which includes correlations in the electron-valence hole complex, the short range exchange of Mn ion with the hole and the electron, the long range electron-hole exchange and the quantum dot anisotropy. A new quantum interference (QI) effect between the electron-hole Coulomb scattering and the scattering by Mn ion has been predicted and observed in the emission spectra as the decrease of emission peak spacing with increasing state of the Mn. This opens the possibility of engineering exciton-Mn spin interaction in quantum dots via quantum interference for quantum memory and information processing applications.

Acknowledgement
The authors thank NRC-CNRS CRP, Canadian Institute for Advanced Research and QuantumWorks for support.

\end{document}